\documentclass[amsmath,twocolumn,showpacs,prb]{revtex4}

\usepackage{graphicx}

 \DeclareMathOperator{\sgn}{sgn}

\begin{document}

\title{Insulating state of granular superconductors in a strong-coupling regime}

\author{I.~S.~Beloborodov,$^1$ Ya.~V.~Fominov,$^{2,1}$ A.~V.~Lopatin,$^1$ and V.~M.~Vinokur$^1$}

\address{$^1$Materials Science Division, Argonne National Laboratory, Argonne, Illinois 60439, USA\\
         $^2$L.~D.~Landau Institute for Theoretical Physics, 119334 Moscow, Russia}

\date{12 July 2006}


\begin{abstract}
We analyze the possibility of the formation of a magnetic-field-induced insulating state in a two-dimensional granular
superconductor with relatively strong intergranular coupling and show that such a state appears in a model with spatial
variations of the single-grain critical magnetic field. This model describes realistic granular samples with the
dispersion in grain sizes and explains the mechanism leading to a giant peak in the magnetoresistance.
\end{abstract}

\maketitle

\section{Introduction}

Recent experiments on superconducting films revealed a giant peak in the low temperature magnetoresistance with the
growth from several times (in polycrystalline samples\cite{Baturina}) to several orders of magnitude (in amorphous
films\cite{Shahar,Gantmakher,Kapitulnik,PHR}), as compared to the value near the superconductor-insulator transition,
and the re-entrant drop upon further increase of magnetic field. The magnetic field-induced insulating phase and the
related superconductor-insulator transition are long known, and extensive theoretical and experimental
efforts\cite{Efetov80,Fisher,Kivelson,Goldman} resulted in a remarkable progress in understanding many features of these
phenomena. Yet the observed giant peak in magnetoresistance is puzzling and poses a challenge for theory. Theoretical
models\cite{SF,FIY} suggested possible development of spatial inhomogeneities in amorphous samples and indicated the
role they may play in formation of the resistivity maximum.\cite{FIY} Consequences of the existence of the
superconductor islands network were explored by the numerical studies in Ref.~\onlinecite{Dubi}; the computed behavior
of magnetoresistance was in a qualitative agreement with the experiment.

\begin{figure}
\centerline{\includegraphics[width=70mm]{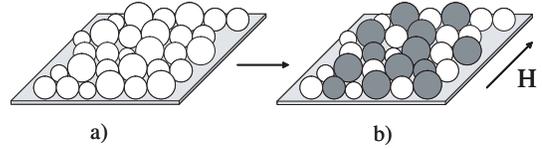}}
 \caption{(a) Superconductor film consisting of grains of slightly different sizes.
(b) The same granular array under applied magnetic field close to the average single-grain critical field. Varying
magnetic field results in a change of the relative concentration of superconducting (white) and normal (gray) grains.}
 \label{array}
\end{figure}

In the present paper we propose and investigate a model for polycrystalline and/or granular films which offers a
mechanism leading to the giant peak in the magnetoresistance. We model the superconducting film in a strong magnetic
field as an array of alternating superconducting (S) and normal (N) granules and demonstrate that the superconducting
insulator (SI) phase appears in the strong coupling regime. In real multi-disperse films the formation of a network of S
and N grains occurs naturally as a consequence of the spatial variations of the single grain critical fields that
inevitably appear due to dispersion of grains' sizes. Magnetic field close to the average single-grain critical field
drives about a half of the grains into a normal state, giving rise to the network of S and N granules, see
Fig.~\ref{array}. Even at a strong intergranular coupling, this array is subject to the Coulomb blockade (similar to a
single SN junction\cite{MG}), leading to the insulating state of the sample. Varying the magnetic field results in the
change of the relative fraction of the S and N grains.

Note that the ``standard'' model of an array of identical S granules\cite{Efetov80} cannot describe the formation of an
insulating state in the regime of a strong intergranular coupling. The charging energy of a single grain is
renormalized\cite{LarkinOvchinnikov} down to the values $\tilde E_c \sim \Delta/ G $ where $\Delta$ is the order
parameter in a single grain and $G$ is the intergranular tunneling conductance. The Josephson coupling $E_J \sim G
\Delta \gg \tilde E_c $ if $ G \gg 1$, and this results in the superconducting rather than the insulating
state.\cite{Kivelson}
%%%%%%%%%%%%%%%%%%%%%%%%%%%%%%%%%%%%%%%%%%%%%%%%%%%%%%%%%%%%%%%%%%%%%%%%%%%%%%%%%
%%%%%%%%%%%%%%%%%%%%%%%%%%%%%%%%%%%%%%%%%%%%%%%%%%%%%%%%%%%%%%%%%%%%%%%%%%%%%%%%%%

We consider the case where most of the grains are so small that the Zeeman effect is at least of the same order as the
orbital one, and, as a result, the transition between the superconducting and normal states within a single grain is of
the first order.\cite{Clogston} Such samples are the good candidates for demonstrating the SI phase since the applied
strong magnetic field $H$ close to the Clogston field $H_0$ will form a strongly inhomogeneous mixture of S and N grains
with the order parameter in the S granules close to that in the absence of the magnetic field, $\Delta_0$.
%%%%%%%%%%%%%%%%%%%%%%%%%%%%%%%%%%%%%%%%%%%%%%%%%%%%%%%%%%%%%%%%%%%%%%%%%%%%%%%%%%%%%%%%%
\begin{figure}
\centerline{\includegraphics[width=65mm]{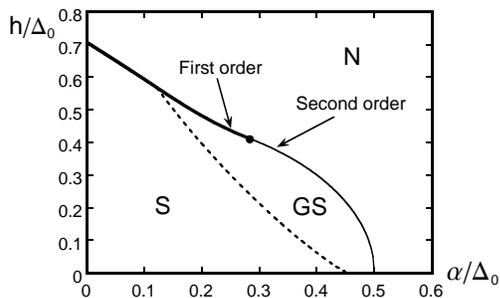}}
 \caption{Phase diagram of a superconducting grain at $T=0$
in coordinates Zeeman energy $h$ vs. pairbreaking parameter $\alpha$. The dashed line separates the gapless (GS) and
gapful (S) superconducting states. First- and second-order phase transition lines are shown by thick and thin solid
lines, respectively.}
 \label{Phase_diagram}
\end{figure}
%%%%%%%%%%%%%%%%%%%%%%%%%%%%%%%%%%%%%%%%%%%%%%%%%%%%%%%%%%%%%%%%%%%%%%%%%%%%%%%%%%%%%%%%%%%
Figure~\ref{Phase_diagram} shows the phase diagram of a single superconductor grain (the derivation is presented below)
drawn in the coordinates Zeeman energy $h=\mu_B H$, where $\mu_B$ is the Bohr magneton, vs. the size-dependent orbital
effect pairbreaking parameter $\alpha$  [in spherical grains\cite{Larkin65} $\alpha = D(e H a/2c)^2/5$, where $a$ is the
grain diameter and $D$ is the bulk diffusion constant]. Varying the grain size one crosses the line of the phase
transition which at strong enough fields is indeed always of the first order.

We now recall a well known result of the percolation theory that in the site problem for the two-dimensional black and
white alternate square (or honeycomb) lattice there exists a range of the relative concentrations of the sites where
simultaneously neither black nor white sites percolate.\cite{Isichenko} In our model, such a ``no-percolation'' state
inevitably leads to transport via sequences of SN junctions; this is an essential ingredient for explaining the SI
phase. To capture this qualitative fact and at the same time to simplify the calculations, we consider a system with the
checkerboard arrangement of the S and N granules.

\subsection{Main results}

We show that the applied strong magnetic field, close to the Clogston field $H_0=\Delta_0/\sqrt{2} \mu_B$, causes the
formation of an insulating state with the gap in the electron excitation spectrum given by
\begin{equation} \label{result1}
\Delta^{(e)} \approx \frac{\Delta_0}{16 \sqrt{2} \pi G} \ln \frac{G E_c}{\Delta_0},
\end{equation}
where $G$ is the tunneling conductance\cite{conductance} (in units of $2e^2/\hbar$) and $E_c$ is the single grain
charging energy. This suggests the conductivity behavior
\begin{equation}
\sigma \sim\exp\left(-\frac{\Delta^{(e)}}T \right).
\end{equation}
Equation~(\ref{result1}) is obtained in the approximation of the nearest neighbor tunneling between the grains.

To verify whether the higher-order tunneling processes can indeed be neglected, we analyze (i)~electron tunneling
between two N grains via a virtual state in an S grain and (ii)~Cooper pair tunneling between S grains via an N grain.
Small amplitudes of these processes ensure the stability of the insulating state with respect to formation of the
metallic and superconducting state, respectively. Analyzing the corresponding corrections we show that the insulating
state is stable as long as $G < G^*$, where
\begin{equation} \label{g^*}
G^* \sim \left( \frac{\Delta_0}\delta \right)^{1/3},
\end{equation}
where $\delta$ is the mean energy level spacing in a single grain. Since the ratio $\Delta_0/\delta$ is large (otherwise
even the mean field approximation for a single grain cannot be used), $G^*$ is larger than the critical value
\begin{equation} \label{G_c}
G_c \approx \frac 1{4\pi} \ln \frac{E_c}\delta,
\end{equation}
that marks the onset of strong Coulomb correlations in the two-dimensional granular array in the absence of
superconductivity.~\cite{ourPRL,BLV05} Thus, for samples with intergranular conductance laying within the interval
\begin{equation}
G_c < G < G^*,
\end{equation}
the insulating state is indeed induced by the local superconducting correlations.

\section{Phase diagram of a superconducting grain}

Now we turn to the analytical derivation of our results.  We begin with the discussion of a single granule of the size
less than the coherence length.  The difference between its thermodynamic potentials in superconducting and normal
states can be written as
\begin{equation}
\Omega = \frac{|\Delta|^2}{\lambda\delta} + \frac{\pi T}\delta \sum_\omega \left[ \alpha f f^\dagger - 2 \tilde\omega
(g- \sgn\omega) - \Delta f^\dagger - \Delta^* f \right] ,
\end{equation}
where $\lambda$ is the electron-phonon coupling constant, $\tilde\omega = \omega-ih$,  $\omega$ is the fermionic
Matsubara frequency, and $g$, $f$, and $f^\dagger$ are the Usadel Green functions\cite{Kopnin} subject to the constraint
$g^2+f f^\dagger=1$. Varying $\Omega$ with respect to $f^\dagger$ under the above constraint results in the
zero-dimensional Usadel equation
\begin{eqnarray}
\label{Usadel_eq} \alpha g f =\Delta g -\tilde\omega f,
\end{eqnarray}
while varying in $\Delta^*$ yields the self-consistency equation
\begin{equation}
\label{BCS} \Delta = \lambda \pi T \sum_\omega f,
\end{equation}
(below we choose $\Delta$ real and $f=f^\dagger$). The second-order phase transition line can be obtained by solving
Eq.\ (\ref{Usadel_eq}) in the first order with respect to $\Delta$ and inserting the resulting $f$ function into Eq.\
(\ref{BCS}). At zero temperature, $T=0$, we arrive at\cite{Beloborodov99}
\begin{equation}
\alpha^2 + h^2 = \frac{\Delta_0^2}4.
\end{equation}
The second-order phase transition line turns into the first-order one (see Fig.~\ref{Phase_diagram}) that can be found
by solving Eqs.\ (\ref{Usadel_eq}) and (\ref{BCS}) under the condition $\Omega=0$. At $\alpha=0$ we reproduce the
Clogston result\cite{Clogston} for the critical field $h_0=\Delta_0/\sqrt{2}$.

The boundary between the gapful (S) and gapless (GS) phases is obtained from the condition that the gap in the electron
spectrum $\Delta_g=0$:\cite{Larkin65,AG}
\begin{equation}
\label{Delta_g} \Delta_g = \left( \Delta^{2/3} - \alpha^{2/3} \right)^{3/2} - h,
\end{equation}
where $\Delta$ has to be found from Eq.\ (\ref{BCS}).  The latter is not affected by the Zeeman term (since the spin
susceptibility in this regime is zero at $T=0$) in the gapful regime and thus assumes a simple form\cite{Larkin65}
\begin{equation}
\ln \frac{\Delta_0}\Delta = \frac{\pi \alpha}{4\Delta}.
\end{equation}
We see that at small enough grain sizes (small $\alpha$) the gapless region does not exist. In what follows we will
focus on the situation where the superconducting grains all have the gap. The validity of this assumption is supported
by the fact that superconducting insulator phase was observed at very high magnetic fields of the order of the Clogston
limit. The gapless region is not favorable for the formation of the SI state.

\section{Array of superconducting grains}

We describe the array of S and N grains by the phase action\cite{AES}
\begin{subequations}
\begin{equation}
\label{action} S = \sum_i \int d\tau \frac{\dot \phi_i^2 (\tau)}{4 E_c} + S_{ns},
\end{equation}
where the first term is the Coulomb part of action with the sum going over all grains. The term $S_{ns}$ describes the
coupling of normal and superconducting grains:
\begin{equation}
\label{actionb} S_{ns} = \frac{\pi G}2 \sum_{\langle ij \rangle} \int d\tau d\tau' \alpha (\tau-\tau') \left( 1 - e^{i
\phi_{{ij}} - i \phi_{ij}'} \right),
\end{equation}
\end{subequations}
where summation goes over the nearest neighboring grains, the phase difference $\phi_{ij} \equiv \phi_{ij}(\tau) =
\phi_i(\tau) -\phi_j(\tau)$ and the kernel
\begin{equation}
\alpha(\tau ) = T^2 \sum_{\omega,\omega'} e^{-i(\omega - \omega')\tau} g_s (\omega) g_n( \omega')
\end{equation}
is expressed through the Usadel Green functions $g_s(\omega)$ and $g_n(\omega)$ of the S and N grains.

In the limit $G \gg G_c$ the term $S_{ns}$ can be simplified by expanding it up to the second order in $\phi_{ij}(\tau)$
and using the local approximation for the kernel $\alpha$ to
\begin{equation}
\label{gauss} S_{ns} = \frac{G g_s'(0)}4 \sum_{\langle ij \rangle} \int d\tau \dot\phi_{ij}^2(\tau),
\end{equation}
where $g_s'(0)$ is the derivative of the function $g$ with respect to $\omega$ taken at $\omega=0$. It can be found from
Eq.\ (\ref{Usadel_eq}), $g_s'(0) = (-\alpha + \Delta /f^3(0))^{-1}$, and in the limiting case that we assume, $h\approx
h_0\gg\alpha$, reduces to
\begin{equation}
\label{gp} g_s'(0) \approx \frac{2\sqrt{2}}{\Delta_0}.
\end{equation}

\begin{figure}
\centerline{\includegraphics[width=75mm]{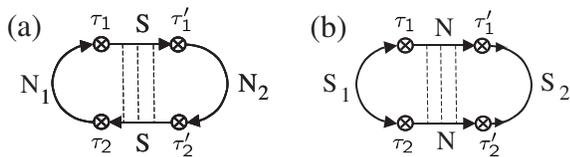}}
 \caption{Diagram (a) represents the coupling of the normal
grains (N$_1$ and N$_2$) via the superconducting grain S. Diagram (b) describes the processes of Cooper pair tunneling
between the superconducting grains (S$_1$ and S$_2$) via the normal grain N. Solid (electron) lines are connected via
dashed lines that represent impurities, forming the zero-dimensional diffuson/Cooperon. Solid lines with two arrows
represent the anomalous Green functions of the superconductors.}
 \label{higher}
\end{figure}

In the Fourier representation we obtain
\begin{equation}
\label{effective} S = \frac {a^2 }{4} \int d\tau \int \frac{d^2 q}{(2\pi)^2} \left[ E_c^{-1} + B (1-E_{\mathbf q})
\right] \bigl| \dot \phi_{\mathbf q} \bigr|^2 ,
\end{equation}
where we have introduced the notations
\begin{equation}
E_{\mathbf q} = \frac 12 \sum_{\mathbf a} \cos \mathbf{qa},
\end{equation}
with $\mathbf a$ being the lattice vectors, and
\begin{equation}
B=8 G g_s'(0).
\end{equation}
Integration over quasimomentum $\mathbf q$ is over the first Brillouin zone. Equation (\ref{effective}) can be
interpreted as the renormalization of the charging energy,
\begin{equation}
\label{effective_energy} E_c\to \tilde E_c (\mathbf q) = \frac 1{E_c^{-1} + B (1-E_{\mathbf q})}.
\end{equation}
The gap in the electron spectrum becomes
\begin{equation}
\Delta^{(e)} = \tilde E_{00} = a^2 \int \frac{d^2 q}{(2\pi)^2} \tilde E_c (\mathbf q),
\end{equation}
where $\tilde E_{00}$ is a diagonal element of the Coulomb interaction matrix $\tilde E_{ij}$ that is given by the
inverse Fourier transform of $\tilde E_c(\mathbf q)$. Within the logarithmic accuracy
\begin{equation}
\Delta^{(e)} = \frac 1{8 \pi G g_s' (0)} \ln \frac{G E_c}{\Delta_0},
\end{equation}
leading to Eq.\ (\ref{result1}) in the case of strong fields where Eq.\ (\ref{gp}) can be used. We see that
$\Delta^{(e)}$ is indeed less than $\Delta_0$ as long as $G \gg G_c$. This justifies the local in time approximation
used in obtaining Eq.\ (\ref{gauss}). We note that in terms of the phase functional there is no difference between the S
and N grains. However, one has to remember that the S grains have the ``big'' single particle gap $\Delta_g$. At small
nonzero temperatures, the electron excitations ``live'' only in the N grains, above the gap $\Delta^{(e)}$. Despite the
superconducting gap, electrons can tunnel between two N grains via a virtual state of an S grain, see diagram in
Fig.~\ref{higher}(a) (this higher-order tunneling process is analyzed below). Therefore the transport has the activation
form with the gap $\Delta^{(e)}$.
%%%%%%%%%%%%%%%%%%%%%%%%%%%%%%%%%%%%%%%%%%%%%%%%%%%%%%%%%%%%%%%%%%%%%%%%%%%%%
%%%%%%%%%%%%%%%%%%%%%%%%%%%%%%%%%%%%%%%%%%%%%%%%%%%%%%%%%%%%%%%%%%%%%%%%%%%%%%%

Now we consider the higher-order tunneling processes so far neglected and find the conditions, at which the contribution
of the diagrams in Fig.~\ref{higher} is not essential.

\subsection{Stability of the superconducting insulator state with respect to the formation of the normal state}

The diagram in Fig.~\ref{higher}(a) should be analyzed in order to check the stability of the SI state with respect to
formation of the normal state. Such tunneling processes lead to an effective coupling of N grains that due to condition
$\Delta^{(e)} \ll \Delta_0 $ can be viewed as an effective tunneling coupling with the
conductance\cite{Averin_Nazarov_2}
\begin{equation} \label{G_n}
G^{(n)}_{ij} = \frac{t_{ij}}4 G^2 g_s'(0) \delta,
\end{equation}
where the dimensionless matrix element $t_{ij}=2$ and $t_{ij}=1$ for nearest and next-to-nearest neighbors in the N
sublattice, respectively. This effective coupling results in the additional term in the phase action
\begin{equation}
S_{nn} = -\frac 1{2\pi} \sum_{ij} G^{(n)}_{ij} \int d\tau d\tau' \frac{e^{i \phi_{ij}(\tau) - i\phi_{ij}(\tau')}}
{(\tau-\tau')^2 } .
\end{equation}
To find the correction to the Coulomb gap $\Delta^{(e)}$ in the first order in $S_{nn}$, we can consider the normal
sublattice of the array only. The problem then becomes equivalent to the calculation of the correction to the Mott gap
in the array of N grains. Using the result of Ref.~\onlinecite{BLV05} we obtain the correction to $\Delta^{(e)}$:
\begin{equation} \label{Gap_correction_C1}
\delta\Delta^{(e)} = - \frac{4\ln 2}{\pi} \sum_i G^{(n)}_{0i} \bigl( \tilde E_{00} - \tilde E_{0i} \bigr).
\end{equation}
Using that $E_c G \gg \Delta_0$ from Eqs. (\ref{effective_energy}) and (\ref{Gap_correction_C1}) we obtain
\begin{equation}
\label{deltacorrection} \delta\Delta ^{(e)} = - \frac{2\ln 2}\pi G \delta.
\end{equation}
Comparing Eqs. (\ref{result1}) and (\ref{deltacorrection}), we conclude that $\delta\Delta ^{(e)} \ll \Delta^{(e)}$ if
\begin{equation}
G \ll \left( \frac{\Delta_0}\delta \right)^{1/2}.
\end{equation}
We note that the effective coupling (\ref{G_n}) also defines the prefactor in the conductivity behavior:
\begin{equation}
\sigma = \frac{4 e^2}\hbar G^2 \delta g_s'(0) \exp\left( -\frac{\Delta^{(e)}}T \right),
\end{equation}
where the function $g_s'(0)$ was defined in Eq.~(\ref{gp}).

\subsection{Stability of the superconducting insulator state with respect to the formation of the superconductor state}

Now we analyze the stability of the insulating state with respect to formation of superconductivity that may be
established due to the Josephson coupling of the S grains, see the diagram in Fig.~\ref{higher}(b). To this end we
introduce an effective Hamiltonian acting on Cooper pairs and defined on the sublattice of the S grains
\begin{equation}
\label{model} \hat H = 4 \sum_{ij} \hat n_i \tilde E_{ij} \hat n_j - \frac J2 \sum_{ij} t_{ij} e^{i \varphi_i - i
\varphi_j}.
\end{equation}
The Hamiltonian is written in terms of the superconducting phase $\varphi$, which is related to the ``normal'' phase
$\phi$ as $\varphi = 2 \phi$. Here $i,j$ belong to the S sublattice, $\hat n= -i\partial /\partial \varphi$ is the
Cooper pair density operator, $J$ is the effective Josephson coupling of S grains via a N grain, and $t_{ij}=2$ and
$t_{ij}=1$ for nearest and next-to-nearest neighbors in the S sublattice, respectively. The insulating state of the
model (\ref{model}) is characterized by the energy gap in the spectrum of Cooper pair excitations.\cite{Sachdev} Due to
the doubled Cooper pair charge in the limit $J=0$ this gap is
\begin{equation}
\Delta^{(s)} = 4\Delta^{(e)}.
\end{equation}
The Josephson coupling $J$ leads to the suppression of $\Delta^{(s)}$ that in the limit of weak coupling can be found
via the straightforward perturbation theory\cite{Sachdev}
\begin{equation}
\delta \Delta^{(s)} = - \frac J2 \sum_i t_{0i} = -6 J.
\end{equation}
The condition $\delta \Delta^{(s)} \ll \Delta^{(s)}$ is satisfied if $J \ll \Delta^{(s)} \sim \Delta_0 /G$. The diagram
in Fig.~\ref{higher}(b) in the case of the large Zeeman energy, $h \approx h_0 \gg\alpha$, gives\cite{coulomb}
\begin{equation}
\label{J} J= \frac{G^2 \delta}4 \ln\left( \frac{\Delta_0^2}{h^2} -1 \right),
\end{equation}
and we arrive at the estimate for the upper boundary for the conductance (\ref{g^*}).\cite{validity} One can verify that
the two-electron (Andreev) tunneling processes between the normal grains via an intermediate superconducting one are
also suppressed under the same condition.

\section{Comparison with experiment}

Now we discuss the relation between our results and experimental data of Refs.~\onlinecite{Baturina}
and~\onlinecite{Shahar}. A note in order is that the proposed model of a granular superconductor in the vicinity of a
superconductor-insulator transition as an normal/superconductor granules alternating chessboard seems to capture
adequately general features of the systems of real morphology and, therefore, allows for the comparison with the
experimental data within its range of applicability. Our results hold for conductances $G > G_c$, where the critical
conductance $G_c$ is given by Eq.~(\ref{G_c}).  This implies that the value of a film conductance $G$ may be less than
unity in real experiment. The resistance of the samples used in experiments of Refs.~\onlinecite{Baturina}
and~\onlinecite{Shahar} was of the order of 20~k$\Omega$, which corresponds to the dimensionless conductance $G \sim
0.1$, i.e., falling on the boundary of applicability of our theory.  Yet one can hope that our results can be used as a
crude estimate.

In Ref.~\onlinecite{Baturina}, the data showing a noticeable peak in magnetoresistance were presented only for the high
resistivity samples. The maximum resistance does not demonstrate the activation dependence on temperature. Apparently
our theory does not apply to this case, presumably due to low intergranular tunneling conductance.

Next, our theory is developed for granular systems and therefore should not apply to homogeneously amorphous samples.
However, it is possible that realistic samples do not exactly correspond to one of the two limiting cases. For example,
amorphous samples can be not uniformly disordered but rather consisting of separate regions (due to fluctuations of
chemical composition or impurity concentration) favoring superconductivity. In this case the system reminds the granular
samples studied in the present paper.

Keeping this possibility in mind, we make estimates for the experiment of Ref.~\onlinecite{Shahar}, where amorphous InO
samples were investigated. We note that in the case of InO films the formation of the ``superconducting insulator''
state was observed in the samples where the values of the resistance at high magnetic fields, while falling into the
metallic domain, were close to the critical resistance that separates insulating and metallic phases at $T \to 0$. Thus,
the estimate of the gap $\Delta^{(e)}$ in Eq.~(\ref{result1}) requires taking $G\approx G_c$. Plugging this into
Eq.~(\ref{result1}), we obtain the gap
\begin{equation} \label{gap}
\Delta^{(e)} = \frac{\Delta_0}{4\sqrt{2}} \frac{\ln(GE_c/\Delta_0)}{\ln(E_c/\delta)}.
\end{equation}
Within the logarithmic accuracy, the ratio of the logarithms is about unity. Finally, in order to compare our estimate
with the experimental findings, we express the superconducting gap $\Delta_0$ in the absence of the magnetic field
through the critical temperature $T_c$. Using the standard relation $\Delta_0=1.76\, T_c$, we arrive at the estimate
\begin{equation}
\Delta^{(e)} = \frac{T_c}{3.2}
\end{equation}
or, equivalently, $T_c/\Delta^{(e)}\approx 3$, which certainly is in a reasonable agreement with the experiment. Indeed,
according to Fig.~4 of Ref.~\onlinecite{Shahar}, the ratio $T_c/\Delta^{(e)}$ depending on the sample varies between 0.6
and 3. Bearing in mind that our model was strictly speaking designed for granular samples, one can hardly expect a
better agreement.

\section{Discussion and conclusions}

Here we comment on the conductivity behavior in the vicinity of the magnetization peak: At lower magnetic fields the
concentration of the superconducting component increases resulting in the formation of the direct Josephson coupling
between the superconducting grains. The Cooper pair gap $\Delta^{(s)}$ decreases and eventually becomes smaller than the
electron gap $\Delta^{(e)}$. At the same time there is no reason for the strong suppression of the electron gap due to
the Josephson coupling. Thus, closer to the insulator to superconductor (IS) transition at finite but low temperatures,
the transport is mediated by the activation of {\it Cooper pairs } rather than by the single electron excitations, hence
$\sigma \sim \exp\left( -\Delta^{(s)}/T \right)$ at $T \ll \Delta^{(s)}$. Thus, we expect that the resistivity within
the IS transition may be described within the effective Bose model as in Ref.~\onlinecite{Fisher}. At the opposite side
of the magnetoresistance peak (i.e., at higher fields) the fraction of the normal component increases leading to the
direct coupling of normal grains. This results\cite{BLV05} in the decrease of the electron gap $\Delta^{(e)}$ and
eventually in the destruction of the insulating state. Note that for the SI state which we studied the spatial
orientation of the applied magnetic field was not essential, however closer to the IS transition one expects it to
become important.

In conclusion, we have investigated formation of an insulating state (corresponding to a giant peak in the
magnetoresistance) in a two-dimensional granular array with alternating superconducting and normal granules. We have
demonstrated that this model is a good representation for realistic granular (or polycrystalline) systems with the grain
size dispersion in a strong magnetic field.

\begin{acknowledgments}
We thank M.~V.\ Feigel'man for bringing our attention to this problem, and T.~I.\ Baturina, K.~B.\ Efetov, and V.~F.\
Gantmakher for useful discussions. This work was supported by the U.~S.~Department of Energy, Office of Science via the
contract No.~W-31-109-ENG-38. Ya.V.F.\ was also supported by the RFBR grant No.~04-02-16348, the RF Presidential Grant
No.~MK-3811.2005.2, the Russian Science Support Foundation, the Dynasty Foundation, the Russian Ministry of Industry,
Science and Technology, the program ``Quantum Macrophysics'' of the RAS, CRDF, and the Russian Ministry of Education.
\end{acknowledgments}


\begin{thebibliography}{99}
\bibitem{coulomb}
The Coulomb interaction on the normal grain can be neglected for $h\gg\tilde E_c\sim\Delta_0/G$. Note, that in the
opposite limit of strong Coulomb interaction, $\Delta_0\gg {\tilde E_c}\gg h$, the result is $J = (G^2\delta/2)
\ln(\Delta/{\tilde E_c})$.
\bibitem{validity}
For $h = h_0 \equiv \Delta_0/\sqrt{2}$, Eq.\ (\ref{J}) gives zero and therefore the result for the Josephson current
should be written more accurately. For small $\alpha$ and $h_1\equiv h_0-h \ll h_0$ we obtain $J =G^2\delta\sqrt{2}(
h_1/\Delta_0 - 2\alpha/3\Delta_0)$.
\end{thebibliography}
\end{document}